# Learning from diversity: "*jati*" fractionalization, social expectations and improved sanitation practices in India

Sania Ashraf[1], Cristina Bicchieri[2], Upasak Das[3], Tanu Gupta[4], Alex Shpenev[5]

**Abstract**

*Prevalence of open defecation is associated with adverse health effects, detrimental not only for the individual but also the community. Therefore, neighborhood characteristics can influence collective progressive behavior such as improved sanitation practices. This paper uses primary data collected from rural and urban areas of Bihar to study the relationship between 'jati' (sub-castes) level fractionalization within the community and toilet ownership and its usage for defecation. The findings indicate a diversity dividend wherein jati fractionalization is found to improve toilet ownership and usage significantly. While exploring the channels, we find social expectations to play an important role, where individuals from diverse communities tend to believe that there is a higher prevalence of toilet usage within the community. To assess the reasons for the existence of these social expectations, we use data from an egocentric network survey on a sub-sample of the households. The findings reveal that in fractionalized communities, the neighbors with whom our respondents interacted are more likely to be from different jatis. They are also more likely to use toilets and approve of its usage due to health reasons. Discussions about toilets are more common among neighbors from fractionalized communities, which underscore the discernible role of social learning. The inferences drawn from the paper have significant implications for community level behavioral change interventions that aim at reducing open defecation.*

[1] University of Pennsylvania, saniashraf@gmail.com
[2] University of Pennsylvania, cb36@upenn.edu
[3] Global Development Institute, University of Manchester, upasak.das@manchester.ac.uk
[4] Indian Statistical Institute, Delhi, Corresponding author, tanug@isid.ac.in
[5] University of Pennsylvania, shpenev@sas.upenn.edu

1.  **Introduction**

Poor sanitation practices that include defecation in open spaces have remained a long-standing issue in India, especially in the rural areas. It has been documented that 48 per cent of the population in India (close to 594 million) practices open defecation (OD).[6] This assumes importance as existing studies indicate that OD can be linked adversely to health measures such as fecal-borne illness, diarrhoea, urinary tract infection symptoms and premature birth (Baker et al. 2017, Baker et al. 2018; Sclar et al., 2018). Apart from the health impact, the economic cost also remains high (Hutton et al., 2008). Additionally, OD can pose significant health hazard to the nearby residents. If neighbors continue to OD, the whole community is at potentially higher health risk compared to those residing in areas with lower prevalence of OD (Cameron et al. 2013). Therefore policies including the Swachh Bharat Abhiyan (SBA), which have been implemented to enhance toilet construction and increase toilet usage for defecation, depend on behavioral change activities that leverage the community or neighborhood and their characteristics to achieve its objectives. In this paper, we examine one key feature of neighborhood- caste diversity among the residents and examine its implications for toilet usage behavior in India. In the process, we assess the role of social expectations and beliefs and then investigate the significance of social learning by exploring the relevant social network characteristics in influencing sanitation behavior.

In the context of India's social structure, caste or "*jati*" (in many Indian languages) is a crucial ingredient of stratification (Deshpande, 2000). Literature has indicated its significance as the basis of social network formation and connection within Indian society (Desai and Dubey, 2011, Munshi, 2019). It is also documented that households from deprived *jatis* such as those from Scheduled Castes (SC) and Other Backward Castes (OBC) groups are less likely to use toilets for defecation (Kumar and Vollmer, 2013; Banerjee et al. 2017). Nevertheless, there has been limited focus in the literature on understanding how social and institutional factors, viewed through the lens of caste, impact sanitation behavior at the community level. India is among the most socially diverse countries in the world with over 3000 different *jatis* (Munshi, 2019) and therefore assessing the effects of caste diversity assumes importance.

---

[6] Please refer https://www.unicef.org/india/campaigns/take-poo-loo (accessed on December 6, 2023)

Existing literature has emphasized the concept of "diversity debit", which suggests that social diversity, measured in terms of ethnicity, is associated with a lower provisioning of public goods. This is often because of a lack of collective action for extracting public goods from local authorities. There are two main reasons for lower collective action: different group preferences (Alesina et al., 1999; Putnam) and the inability to impose social sanctions (Miguel and Gugerty, 2005). These factors are consistent with the diversity debit hypothesis and suggest the multiple ways in which diversity can negatively impact the provision of public goods. Furthermore, in ethnically diverse communities, a lower level of social cohesion, trust and altruistic cooperation in ethnically diverse communities can exacerbate these challenges (Habyarimana et al. 2007, Putnam, 2007).

Previous research has predominantly documented significant gains from lower caste diversity in India in public good provisioning (Balasubramaniam et al. 2014). This suggests that lower diversity within the community may be conducive for improved sanitation behavior. In areas with less diversity, it is possible that higher toilet usage among the dominant *jatis* may influence those from the non-dominant jatis within the same community to adopt better sanitation practice. This influence can be facilitated through coordinated actions especially when open defecation practice is socially conditioned. Additionally, the higher likelihood of imposing social sanctions within less diverse communities can make it easier to sustain a desired behavior, thereby facilitating toilet usage in such areas.

Contrarily, in some contexts, it is possible that caste diversity serves as a catalyst for improved sanitation practices. Ethnic diversity can foster cultural exchange and knowledge sharing related to sanitation and general health practices. This diversity can allow people from different caste backgrounds to learn from each other's experiences and adopt positive health behaviors that are effective in different communities, thereby encouraging social learning. Increased interpersonal contact can reduce the prevalent prejudices between groups and facilitate knowledge exchange among them (Allport et al., 1954). Therefore, higher inter-caste contact can challenge the existing prevalence of OD practice, form positive social expectations and beliefs, and help erode the regressive norms surrounding poor sanitation practices. This, in turn, can reduce prejudices surrounding toilet usage and increase awareness of its associated benefits (Fielding, 2022).

In this paper, we examine whether there exists a *jati* level diversity debit on improved sanitation practices. More specifically, we assess the effects of fractionalization or diversity in terms of sub-castes (or *jati fractionalization*) at the neighborhood on two indicators: ownership of a private toilet in the household and its usage for defecation purpose.[7,8] We also assess the variation in the existing social beliefs and expectations surrounding sanitation behavior to study the mechanisms that explain the effects we observe. We utilize primary survey data of 2533 households collected in 2018 from eight districts across rural, semi-urban and urban areas in the state of Bihar, India. To gain deeper insights into neighborhood interactions, beliefs and behaviors that explain the main observed impact of diversity, we also incorporated data from an egocentric social network survey conducted on a subset of the sample in 2017.

Our findings indicate a significant and a positive effect of *jati fractionalization* (caste diversity) on improved sanitation behavior. In addition to an increase in the likelihood of ownership of a private toilet, we also observe higher toilet usage for defecation among households, located in a more fractionalized community. To ensure that the effects are not confounded by unobserved heterogeneity, we use bias adjusted treatment effects (Altonji et al., 2005). These adjusted effects suggest that our estimates are unlikely to be zero, even when considering potential unobservable variables that might influence the outcome. We also estimate two stage least square regressions using fractionalization of the broader socio-religious groups as instrumental variable (IV) to adjust for the possible unobserved heterogeneity. The results indicate higher toilet ownership and usage in caste fractionalized areas. We also relax the assumption of exogeneity of the IVs through "plausibly exogenous" regressions and found that our inferences hold, which lends confidence about the observed causal effects.

To study the mechanism, we assess the channels/ factors behind increased toilet usage in fractionalized communities by exploring the role of personal beliefs and/ or social beliefs and expectations surrounding improved sanitation behavior. Broadly, we argue that the higher toilet usage observed in fractionalized communities can be a collective behavior, which might be independent, driven only by factual beliefs or Personal Normative Beliefs (PNB) around toilet usage. Alternatively, it may also be interdependent, influenced by how individuals in their relevant community or neighborhood or reference group perceive the prevalence of toilet usage (Empirical Expectations or EE), and the approval of such behavior within their community (Normative Expectations or NE). Despite observing systematically positive personal beliefs and perceptions about community approval on toilet usage in *jati* fractionalized communities, we do not find any evidence to indicate that PNB or NE has been critical in raising toilet usage. Rather, an increase in EE or belief about prevalence of toilet usage in these *jati* diverse communities emerged as a key factor in raising toilet usage levels and ownership. To this end, the chances of having biased or false but positive beliefs about improved sanitation behavior are also higher in these diverse areas and are likely channels that influence toilet usage positively.

---

[7] Each broad caste group (SC, OBC and upper castes) comprises many sub-castes (or *jatis*), historically based on division of jobs (Deshpande, 2011)
[8] We use the terms fractionalization and diversity interchangeably.

Why is there a systematically higher level of social expectations (EE) surrounding toilet usage and access in these *jati* diverse areas? Why do people staying in fractionalized localities think others in their community are more likely to use toilets for defecation, even beyond the actual prevalence? To answer this, we use ego-centric network data that we collect to understand toilet access and usage among the individuals our respondents speak to or depend on for personal help. This data allows us to map the connections or networks of the respondent and gather information about these networks from the respondent. Analysis using this data indicates that the proportion of networks within the locality that use toilets is higher if the respondents reside in more fractionalized communities. Furthermore, the networks in the neighborhood of the respondents from fractionalized communities, with whom they interact, are more likely to endorse the use toilets for health-related reasons, such as reducing pollution or spread of diseases. The nature of interaction is also related to the toilet models or the availability of masons. Importantly these networks within fractionalized communities are less likely to be from the same caste as the respondent. These results are indicative of the effects of interpersonal interactions and social learning on toilet usage among the residents in caste-diverse neighborhoods. Notably, we are presence of Muslims, supply side government and private sanitation interventions, and functionality of Self Help Groups (SHG), all of which might be disproportionately higher in diverse areas.

The paper makes a contribution in four distinct ways. First, it challenges the prevailing notion of a "diversity debit" often associated with public good provisioning, economic growth, health and educational indicators (Easterly and Levine, 1997; Khwaja, 2009; Churchill et al., 2019; Churchill and Smyth, 2020). Studies based in India also document similar findings (Banerjee and Somanathan, 2007; Balasubramaniam et al., 2014). This paper provides empirical evidence that supports a positive association between social diversity and welfare outcomes and contributes to its growing literature (Gisselquist, 2016; Dinku et al., 2019; Montalvo and Reynal-Querol, 2021; Fielding, 2022).

Second, literature on diversity and healthcare has looked at its effects on indicators like health-related facilities and infrastructure, mortality and life expectancy, immunization, anthropometric measures and prevalence of disease (Banerjee and Somanathan, 2007; Churchill et al., 2017; Dinku et al., 2019; Fielding, 2022). The paper looks at a different dimension of healthcare which is improved sanitation practices. Important to note is we not only consider health amenities through ownership of a private toilet but also the behavior associated with it, which is toilet usage for defecation purposes. Differentiating the two is important in the context of sanitation in India as there exists a substantial prevalence of OD even among those with access to toilets (Bicchieri et al., 2018) and a considerable slippage in toilet usage over time (Coffey et al., 2017; Abebe and Tucho, 2020). Relatedly, in our sample, about 6.6 percent of the respondents reported not having used a toilet last time they had to defecate despite having access to a toilet. 25 percent of those having access to a toilet did not use it exclusively for seven days prior to the survey.

Thirdly, we use *jati* level data to assess the caste diversity effects, unlike previous studies that used the broad caste categories of SC, ST, OBC and upper castes. This distinction is crucial because there is significant heterogeneity among the *jatis* with a broad caste group in terms of the social hierarchy. For example, certain *jatis* like *Bhangi, Chamar*, and *Chuhra* are the more disadvantaged ones, while *Julaha*, *Meghwal*, *Pasi, despite being in the SC category,* are relatively less disadvantaged. Similarly, *Baniya* and *Madeshiya* are the weaker *jatis* while the *Chaudhari* and *Rajputs* are the stronger ones with the upper castes. Understanding these distinctions between these *jatis* is imperative because trust and social interaction primarily occur within these *jatis* and to a lesser extent, across them.

Fourth, we explore the implications of socially conditioned beliefs and expectations surrounding toilet usage behavior in caste fragmented communities, a mechanism that has not been covered in detail in the existing literature on diversity dividend. We find households from caste diverse communities are likely to own and use a toilet. This inclination is reinforced through EE and false positive beliefs about the prevalence of toilet usage within the community. We further utilize information collected from egocentric social networks to assess the implications caste diversity may have on social learning and relate that to social expectations. Our findings present compelling evidence of a mechanism to explain the positive effects of social diversity- social learning leading to higher EE and thereby potentially creating a descriptive norm (Bicchieri, 2005). One important dimension to add here is that our paper contributes to laying down a brief application of using egocentric network data in studying social dynamics and behavior, which has been less utilized in Economics literature. In terms of policy implications, we argue if social interactions can be strengthened in diverse areas, it may potentially lead to pro-social outcomes. On policies surrounding sanitation practices, our findings provide support to behavior change interventions that leverage social norms in one's communities to influence improved toilet usage (Harter et al. 2019). Norms are strengthened when people see others near or like them (in-group) do or approve a behavior (Deutsch and Gerard, 1955).

The paper is structured as follows. Section 2 describes the relationship between caste diversity and social expectations. Section 3 describes the data and descriptive statistics. Section 4 lays down the empirical strategy. The results are explained in Section 5 and the last section concludes.

## 2. Caste diversity and social expectations

In this paper, we study the implications of caste diversity on sanitation behavior and explore the role of social expectations, which form the basis for understanding the mediating channels. The idea flows from the Social Norms Theory conceptualized in Bicchieri (2005) and Bicchieri (2016). The broader argument is as follows. Some collective behaviors may be completely independent as "they are purely determined by economic or natural reasons" (Bicchieri, 2016; pg. 2). These independent behaviors are driven by two categories of beliefs. The first one is factual beliefs, which indicate beliefs about how the world is. For example, one may not bribe to get things done because she believes bribing degrades the society. The second one is the Personal Normative Beliefs (PNB), which indicate beliefs about what it should be. As an instance, someone may not bribe because she personally thinks people should not bribe to get things done.

However, when the preferences or actions of others influence one's own behavior, it becomes interdependent. In other words, people may have a social preference to follow certain behaviors not only because they believe it is commonly practiced within their close networks, but also because people approve it. Therefore, one's preferences can be driven by her social expectations such as what others who matter to them do and/ or believe in. Notably, in many contexts, these expectations might be upward or downward biased when compared to the actual prevalence of that practice or the beliefs surrounding it. Yet these false and biased beliefs may be perceived to be correct and unbiased, thereby continuing to have a significant bearing in shaping up one's preference or behavior (Bicchieri, 2005; Reid et al., 2010; Sunstein, 1996; Vaitla et al., 2017).

Here it is important to distinguish between two key elements of social expectations when we discuss interdependent behavior: EE and NE. While EE refers to one's expectation about what others around them or in close networks do, NE is the expectation about what others in this network think one should do. Bicchieri (2005) argues that certain behaviors such as fashions, rules of etiquette and conventions, are often followed because one prefers to imitate or coordinate with common behaviors that are expected to occur and hence driven by EE. Behavior like tipping in a restaurant, on the other hand, might be followed not only because other patrons will leave a tip but also, they think that others would disapprove of not leaving a gratuity, and hence also driven by NE. Therefore, interdependent behaviors that emanate from social preferences can be influenced by EE, NE or a combination of the two. If someone engages in a certain behavior because she thinks others in her reference network also engage in that behavior (EE), then the behavior is said to be driven by descriptive norms. However, in addition if it is driven by her perception about the approval of behavior within her network, then social norm is said to have driven the behavior (Bicchieri, 2006).

In the context of this paper, we link *jati* diversity in the neighborhood with social expectations. If higher diversity hampers the level of socialization and cohesion within the community, the relevance of social expectations may reduce, thereby EE and NE would be less likely to influence behavior. In a more homogenous neighborhood, preferences are likely to be similar and people often do not feel differentiated. With higher social capital, and trust, behavior potentially becomes more interdependent and likely to be driven by social expectations about the actions and beliefs of others in the neighborhood. However, one critical feature of diverse communities that has been documented is higher levels of social learning among the residents of the communities. Here, when individuals interact with others, who are from diverse backgrounds, they can learn about different beliefs and progressive practices from each other, thereby enhancing their knowledge (Gisselquist et al., 2016; Dinku et al., 2019). The inter-personal contact can reduce prejudices among the groups and in fact can improve trust and foster empathy through which there might be an erosion of the existing regressive norms and acceptance of progressive behavior and norms (Allport et al., 1954; Scacco and Warren, 2018; Finseraas et al., 2019; Fielding, 2022). Under such conditions, social preferences may gain prominence, thereby, EE and/or NE can potentially influence one's actions and behavior, making it interdependent. In this paper, apart from assessing the implications of caste fractionalization on sanitation behavior, we explore the role of social expectation and gains from social learning that can be associated with a socially diverse region.

## 3. Data and Descriptive Statistics

We use primary survey data from the Longitudinal Evaluation of Networks and Norms Study (LENNS) conducted from 2017 to 2018 in eight districts of the state of Bihar, India. This formative study aimed to illuminate the social determinants associated with open defecation and toilet use behavior. The survey design consists of a randomly chosen sample of individuals aged between 16 and 65 years from thirty sampling units coming from eight districts. These sampling units were selected on the basis of three different types of geographic regions, including six registered slums in highly urban Municipal Corporations, eighteen census wards from semi-urban *Nagar Panchayats* (Town Councils), and six *Gram Panchayats* (Village Councils). These units are the Primary Sampling Unit (PSU) in our case. In each sampling unit, a complete listing of the entire age-eligible population was conducted, and eligible respondents were randomly selected for the survey. Data on individual and household characteristics, including age, education, gender, religion, and household asset composition, along with toilet use behavior, toilet ownership and caste measures, were collected in the survey. It also gathered information on factual beliefs, PNB, EE and NE, which we use to examine the role of social expectations as mechanism. The total number of respondents in the survey is 2533.

Additionally, we make use of information collected from an egocentric network survey through which we obtain data about individuals with whom respondents interact about the construction of toilets and their usage. An egocentric network survey focuses on the individual (known as the egos) and explores her social networks with people with whom she has direct social connections with, also known as the alters (Carrasco et al., 2008). Typically, information on the ego's (respondent) social network is collected that include the size and composition of the network, the frequency and nature of interactions with different alters on specific behavior. We conducted an egocentric survey on a subset of the main sample in 2017 (1501 respondents). The survey gathered information on interaction of the egos with their alters about toilet construction, its usage and cleaning and exploring their characteristics, including the place of residence. More details about the egocentric survey are presented in Section 5. Literature has documented how the egocentric network approach has been utilized to study a range of health and pro-social behavior (O'malley et al., 2012; Cao et al., 2020)

*Outcome variable*

Our primary outcome variables revolve around toilet ownership and defecation practices. The survey asks questions to respondents about ownership of toilets and their toilet usage behavior for three different recall periods. In particular, it asks each respondent if their household owns a toilet or not to measure toilet ownership. A dummy variable is created, which takes the value 1 if the respondent's household owns a toilet (either owned solely or shared with other households) and 0 otherwise. It further asks whether they used the toilet or went in the open, (a) the last time they defecated; (b) the last three times they defecated; and (c) the last week they defecated. For instance, to measure the respondent's toilet use behavior for the last time they defecated, the survey asked the following question: *"Some people defecate in the open; some people use a toilet, where did you defecate the last time, you had to?"* A dummy variable has been created which takes the value 1 if the respondent had used any of the toilets and 0 if the respondent has defecated in the open. Similarly, we create two other measures of toilet use behavior using other two recall periods- last three times and last seven days. The exact questions and variable construction are outlined in Appendix A. The summary statistics of the variables constructed are given in Table 1. We find only 59.2 percent of the respondents reported of having used the toilet last time when they defecated and 58.7 percent of the respondent's own toilet. The other two measures of toilet use show that around 50 percent of the respondents are consistently using toilets for defecation purpose.

*Jati Fractionalization Index*

The survey collects information on caste groups to which respondents identified with, which includes four broad categories: Scheduled Caste, Scheduled Tribe, Other Backward Caste (OBC), or Others (including Generals). Additionally, the survey also collects information on the specific sub-caste or *jati* of these households. In particular, respondents in the survey were asked about the *jati* with which they identify. In India, the caste system at ground level is governed by *jati*, with deep social cleavages governing social and economic interaction between these groups. The term *jati* is related to the idea of lineage or kinship group. These *jatis* represent different layers of hierarchy that exist in society, which get enveloped into broader caste categories. There are perhaps more than 3000 *jatis* in India (Dumont, 1980; Singh, 1992).

These *jatis* can vary significantly in terms of social, economic, and political standing. At the top of the caste hierarchy traditionally lie the *Brahmins*, who are revered for their association with knowledge and spirituality, while at the bottom are the *Dalits* like *Chamar* and *Bhangi*, historically marginalized and subjected to social discrimination. In between, there is a vast spectrum of *jatis*, each with its own unique position. These differences across *jatis* are marked by varying occupational roles, social privileges, and rituals. Marriages are often restricted to within the same *jati*, and the *jati* system significantly influences everyday life, including dietary practices and social interactions. Another feature of *jati* system is social segregation. Individuals of different *jatis* do not intermingle socially, leading to separate neighborhoods, places of worship, and even separate utensils for eating. *Jatis* also serve as centers of identity, mutual support, and cultural heritage for their members. Measuring caste diversity at the *jati* level can provide a more comprehensive and accurate understanding of the social and cultural landscape in India.

We measure caste diversity at the primary sampling unit (PSU). As popular in the diversity literature, we construct fractionalization index to measure caste diversity using the following formula:

$$Jati\ Fractionalization_j\ =\ 1 - \sum_i s_{ij}^2 \qquad (1)$$

where $s_{ij}^2$ is the square of the population share of *i*th *jati* in PSU *j*. We also interchangeably called this as *jati* diversity index. This index measures the probability that two individuals drawn from a given PSU *j* will have the same *jati*. Here we use PSU for defining neighborhood boundaries, which is slum for urban areas, census ward for semi-urban and gram panchayat for rural region. These PSUs are administrative areas nested within districts, which means that we capture the caste diversity at the lowest level of geographic disaggregation.

Table 1 shows the mean value of the *jati* fractionalization *or jati* diversity index, which is equal to 0.714, meaning that the neighborhoods in Bihar are relatively heterogeneous in terms of caste diversity. The sampling distribution of *jati* fractionalization index is additionally presented in Appendix Figure B1, which depicts the fair amount of variation in fractionalization index.

*Social Expectations*

To measure individual-level EE, we ask each respondent the following question *"Out of ten members of your community, how many do you think used a latrine the last time they needed to defecate?"* To measure individual-level NE, we ask each respondent the following question *"Out of ten members of your community, how many do you think believe that it is wrong to defecate in the open?"* Note that these expectations measure the individual-level perceived beliefs on the prevalence of toilet use in their community. The response to these individual-level beliefs is recorded on the scale of 0-10 and we linearly transform them on a scale of 0-1 to keep measure in terms of proportions.

We also quantify the overestimation bias in beliefs regarding the perceived prevalence of toilet use, which is defined as the discrepancy between the individual-level perceived prevalence and community-level actual prevalence of toilet usage. We calculate this overestimated bias for both empirical and normative beliefs. To do so, we first calculate the community-level prevalence of toilet usage, which is proxied by calculating the share of toilet users across each PSU.[9] Then we subtract the individual-level perceived prevalence from community-level prevalence on toilet use behavior. The bias in the perceived prevalence ranges from -1 to 1, where negative scores represent the underestimation of toilet use prevalence, 0 represents the accurate perception and positive score represents overestimation of toilet use prevalence. Next, we transform these scores into a binary variable which takes value 1 for positive scores and 0 otherwise.

---

[9] We use the measure of toilet use behavior for the last time they defecated to calculate the share of toilet users in each PSU.

We use two questions to elicit the factual beliefs and PNB using the following questions:

Factual beliefs: *"Do you personally believe there are any bad health effects of open defecation for you?"*

PNB: *"Society may think it is right or wrong to defecate in the open. Do you personally think it is right, neither right nor wrong, or wrong for someone to defecate in the open?"*

For factual beliefs, we assign the value of 1 if the respondent reply is a "yes" and 0 if "no". For PNB, the value of 1 is assigned if the response is "wrong" and 0 otherwise.

Table 1 shows that respondents think more than 60 percent of people in their community have used the toilets for defecation (EE) and 85 percent of people believe that it is wrong to defecate (NE). Respondent's beliefs on NE are much higher than EE. If we look EE and NE in terms of bias, we find that 56 percent of respondents overestimate the toilet usage in their community, implying they have false positive beliefs about the actual prevalence of toilet use for both EE and NE. Moreover, 87 percent of respondents believe that there are bad effects of open defecation (factual beliefs) and 96 percent of respondents believe that it is wrong to defecate in the open (PNB).

## 4. Empirical Model

To examine the relationship between toilet use, ownership and caste diversity, we estimate the following model:

$$Y_{ijd} = \beta_1 \, Frac_{jd} + X'_{ijd}\, \boldsymbol{\delta_1} + M'_{jd}\, \boldsymbol{\delta_2} + \alpha_d + \epsilon_{ijd} \qquad (2)$$

where $Y_{ijd}$ measure toilet use and toilet ownership of respondent $i$ residing in PSU $j$ of district $d$. $Frac_{jd}$ measures the *jati* diversity using Equation (1). $X'_{ijd}$ is a vector of individual and household controls. The individual-level variables include age, gender, and education qualification of the individual, while the household-level variables include religion, presence of ration card, asset ownership and household economic status. $M'_{jd}$ includes the set of variables at the PSU level, namely the average number of households with BPL ration card, proportion of people of go outside to fetch water and the proportion of literate women. The summary statistics of these variables are presented in Table 1. We include district fixed effects ($\alpha_d$) to control for the time-invariant unobserved heterogeneity at the district level. Given the sample is distributed across three different kinds of geographic regions (rural, slum and urban areas), we also include region fixed effects to control for region-specific unobserved effects. The standard errors are clustered at PSU level to control for heteroscedasticity and correlation among observations within the same PSU. After controlling for all covariates, the coefficient $\beta_1$ will give unbiased estimates of *jati* diversity on toilet use and ownership.

One major concern about viewing these results from a causal lens is the endogeneity of *jati* fragmentation, through which unobserved heterogeneity can confound the regression estimations presented above. One potential source of endogeneity might be through migration of households in areas with lower diversity, owing to evidence provided by existing literature that indicates how *jati*-based homophily can lead to residential segregation (Curtler et al. 1999). However, it has been found that spatial mobility in India at the household level is low (Munshi and Rosenzweig 2009; Kone et al. 2018). While we did not collect data on household relocation in our survey, the conversations with the officials also confirmed about limited relocation of households in the respective villages. Nevertheless, to account for the possibility of any unobserved heterogeneity confounding our results, we first use bias-adjusted treatment effect estimates to assess the variation in the estimates in the presence of unobservables. The method assumes that the selection on observables is proportional to the selection on unobservables and the extent of this is determined by a hypothetical parameter δ (Altonji et al. 2005; Oster, 2019). Another parameter ($R^2_{max}$), which is the R-squared value of a hypothetical regression that incorporates all potential observable and unobserved variables into the regression. Using data from a wide range of randomized experiments, Oster (2019) proposes the value of $R^2_{max}$ to be 1.3* $R^2$, where $R^2$ is the R-squared value of the model with all observed control variables. Using the values of $R^2$ derived from the naïve regressions and then $R^2_{max}$, the value of δ is estimated. Of note is the fact that with a comprehensible set of control variables, the effect of unobservables is assumed not to be more important than that of observables and therefore the bound on δ that is considered should lie between [-1,1] if unobservables are influential (Altonji et al. 2005, Rathore and Das 2021; Bevis et al. 2023).

Next, we use two-stage least squares (2SLS) regressions considering fractionalization of broader socio-religious groups as the IV. To categorize the socio-religious groups, we classify the households based on broader caste categories and religion. These groups are as follows: General-Muslim, General-Hindu, General-Others, OBC-Muslim, OBC-Hindu, OBC-Others, Others-Muslim, Others-Hindu, and Others-Others.[10] We use the proportion of people belonging to each of these groups to construct fractionalization index as elucidated in Equation 1 and use it as IV at the PSU level. The rationale for using this indicator as IV is as follows. A number of *jatis* come under the ambit of each of the broader socio-religious groups as defined above. Therefore, it is likely that if the diversity of these groups within the PSU is high, the *jati* level fractionalization will also be high. Hence, PSU level socio-religious fractionalization can show substantial correlation with the endogenous primary variable of interest, which is *jati* fractionalization. On exogeneity of the IV, once *jati* fractionalization is taken into account, the role of the broader socio-religious categories in influencing behavior or public good provisioning becomes irrelevant. In other words, the decision and behavior of an individual can potentially depend on the *jati* level composition in her locality and not on the composition of broader caste-religious categories. This is also based on the literature on caste, which documents that the network and social cohesion is more often based on the specific *jatis* and not the broad categories. For example, a *Brahmin* is likely to form stronger social ties with another *Brahmin* and not with a Rajput though both belong to the Hindu-General caste group. Therefore, we argue, after accounting for *jati* fractionalization, the association with the socio-religious diversity using broader caste groups in the PSU with the unobserved error terms affecting toilet ownership and usage is likely to be very limited through other channels. Hence, independently, it is unlikely that socio-religious diversity can influence the outcome variables of interest through channels other than *jati* diversity. We test this empirically by regressing the outcome variables on jati fractionalization along with the IV and other covariates. Appendix Table B1 (columns 1-4), which presents the results, shows that the correlation of the IV with the outcome variables is negligible once the *jati* fractionalization is incorporated in the model.

### 5. Results

*Does Jati diversity have any influence on toilet use behavior and ownership?*

---

[10] Others include Schedule Caste (SC) and Schedule Tribe (ST).

We start by reporting the results for four outcome variables for toilet use and toilet ownership. As indicated in Table 1, we use three indicators of toilet usage. TOILETUSE1 is an indicator variable which captures whether the respondent used a toilet the last time he or she had to defecate. Similarly, TOILETUSE2 and TOILETUSE3 are indicator variables that measure whether the individual used the toilet exclusively, in the last three times and the last week when he/she had to defecate respectively. These variables measure consistency in individual's behavior of toilet use. For toilet ownership, we estimate whether the respondent owns a private latrine.

The estimates of the regressions that assess how *jati* diversity is related with toilet usage and ownership are presented in Table 2. We use two specifications for each of these four outcome variables. In the first specification, we include all individual and household level characteristics and in the second, which is also the preferred one, we include those at the PSU level as well. The first column for each outcome variable provides the estimates with the first specification and the second one provides those with the second specification. The findings from the regressions indicate that respondents residing in area with higher *jati* fractionalization are more likely to own toilets and use them for defecation purpose. The regression estimates suggests that one standard deviation increase in *jati* diversity increases the probability of owning private toilet by 15 percentage points on average. It is also found that a similar increase in *jati* fractionalization is associated with 20.1 percentage points increase in toilet usage the last time one had to defecate before the survey. Notably, we find a statistically significant increase by about 15.6-15.3 percentage points when toilet usage in the last three times or seven days is considered, indicating that diversity in *jati* is also associated with consistent usage of toilets for defecation purpose. Therefore, to sum up, our findings underscore a discernible improvement in sanitation practice in areas with higher *jati* fractionalization, providing evidence in favor of the diversity dividend hypothesis.

As discussed in Section 4, we the estimate bias adjusted treatment effects proposed by Altonji et al. (2005) and Oster (2019) using the *psacalc* command in STATA. The values of $\delta$ are found to be 2.1, 1.8, 1.2, 2.2 respectively for toilet ownership and its usage the last time one had to defecate, the last three times and in the last seven days. These values of $\delta$, as one can observe lie well beyond the acceptable range of [-1,1] and therefore the chances of the impact estimate using naïve regressions falling to zero are highly unlikely even if we assume that the presence of substantial unobservable confounding our estimates.

Next as discussed, we run 2SLS regressions using fractionalization of broader socio-religious groups as the IV. Table 3 presents the estimates from the 2SLS regressions using proportion of socio-religious group at the PSU level as the IV. The F-statistic of the IV is 22.2, which is well above the threshold of 10, thereby showing that their association with the *jati* diversity measure is strong (Stock and Yogo, 2002).[11] The findings indicate qualitatively similar findings: *jati* fractionalization influences toilet ownership and usage significantly.

Despite observing negligible correlation of the IV with the unobserved error terms, one might still raise questions on the exclusion restriction and therefore its validity. To counter this, we use the "plausibly exogenous" inference proposed by Conley et al. (2012) that allow for construction of IV bounds with weaker assumptions on the exclusion restriction. Here we use the approach followed by a number of literatures to generate the bounds (McArthur and McCord, 2017; Azar et al. 2022, Biswas and Das, 2022). In this approach, we consider the following equation:

$$Y_{ijd} = \beta_1 \, Frac_{jd} + \gamma Z_{jd} + X'_{ijd} \, \delta_1 + M'_{jd} \, \delta_2 + \alpha_d + \epsilon_{ijd} \qquad (3)$$

Here $Z_{jd}$ is the IV defined at the PSU level, $j$ which is socio-religious caste fractionalization. The other notations represent what has already been mentioned in Equation 2. The regressions presented above assume the IV to be fully exogenous and hence $\gamma = 0$. However, if it is only not fully exogenous, then $\gamma$ might be close to zero but not exactly zero. In such a condition, we consider the IV to be "plausibly exogenous", where the effect of the endogenous primary variable of interest on the outcome variable can be consistently estimated for a range of $\gamma$. Following Azar et al. (2020), we first estimate the upper bound of $\hat{\gamma}$ by regressing the toilet ownership and usage on the IV along with other covariates but not *jati* fractionalization. Next, we estimate the bounds of the impact of fractionalization on sanitation indicators of practices using the values of $\gamma$ in the range of 0 to $\hat{\gamma}$. We also report the maximum value of of $\gamma$ (given by $\gamma_{max}$) for which the estimated bound of $\gamma$ is strictly positive. Using the values of $\gamma_{max}$, we find that the estimated impact of fractionalization on improved sanitation practice would remain positive even when the direct effect of the IV is 73-80 percent of that from the reduced form (Appendix Table B2).

---

[11] The first-stage estimates of 2SLS regressions are given in the last column of Appendix Table B1.

The estimates using different variants of the outcome variable measuring toilet usage, model specifications, IV regressions with "plausibly exogenous" assumptions and bias adjusted treatment effects that we present indicate the findings are highly robust. In addition to these, we also estimate the regressions presented in Appendix Table B3 using probit model instead of LPM. Further, we use access to toilets for defecation instead of ownership, which also includes households who use toilets owned by others as an outcome variable. In all these cases, the results remain qualitatively similar: individuals from areas with diversity in *jati* are more likely on average to have access to toilets and use them for defecation purpose (Appendix Table B3). Additionally, we estimate the wild bootstrapped standard errors to correct for the problem with lower number of clusters (Roodman et al., 2019). The p-values of these bootstrapped standard errors are presented in Appendix Table B4 and we find that our main results remain statistically significant emphasizing the robustness of our finding.

Importantly, we also use data from Census 2011 at the village and ward levels to construct caste-fragmentation index.[12] This allows us to create our main variable of interest using an alternate data which is not based on sample survey but on the actual population. The regression estimates using this variable also indicate qualitatively similar findings: in areas with higher caste fragmentation, we observe higher likelihood of toilet ownership and its usage (Appendix table B5). However, it should be noted that Census does not provide information on *jati* but only that on broader caste groups of SC, ST, OBC and General. Therefore, we use this data only for robustness check and not for the main analysis where we exploit the variation in *jati* level fractionalization.

*The role of social expectations*

In this subsection, we explore the role of personal beliefs and/ or social beliefs and expectations as mechanisms through which *jati* diversity can influence the collective behavior of toilet usage and ownership as discussed in Section 2. In particular, we study the implications of *jati* fractionalization on the nature of this collective behavior- whether it is driven independently or if it is interdependent on each other. For this, we use the collected data on factual beliefs and PNB to understand if the nature of toilet usage is independent. Next, we also use data on social beliefs and expectations namely EE (one's beliefs on the prevalence of toilet use in community) and NE (one's beliefs on what people approve for defecation behavior). This allows us to understand the interdependent nature of sanitation behavior.

---

[12] More information on Census 2011 can be obtained from https://censusindia.gov.in/census.website/ (accessed on December 4, 2023)

We consider the indicators for factual belief and PNB as outlined in Section 3 as the outcome variable and regress them on *jati* fractionalization measure and other covariates. Table 4 presents the results from these regressions. While the association with PNB is found to be positive and significant, that on factual beliefs remains statistically indistinguishable from zero (column 1 and column 2). This implies that *jati* level fractionalization may have a role to play in raising the PNB about improved sanitation practices among people, though no such influence on factual beliefs about the same is observed.

To explore the relevance of social beliefs and expectations, we use EE and NE surrounding toilet usage among the respondents as the outcome variables and regress them on *jati* fractionalization. From Table 4, which presents the regression results, we find that individuals residing in more *jati* diverse regions have disproportionately higher EE on toilet usage (column 3). In other words, people in more diverse communities think that the prevalence of toilet use is systematically higher in their community. Regression estimates of NE on fractionalization also indicate that the diverse areas have higher perceived approval of toilet usage in the community (column 5).

One key aspect of EE and NE is the accuracy of the beliefs, thereby leading to a bias in perception, which can play a major role in influencing individual behavior. This directly links with the literature which documents about the subjective nature of the perception of an individual that also may not coincide with the actual frequency of that behavior within the community (Suls et al., 1988, Kruglanski, 1989). If individuals tend to have a higher perceived prevalence about the community's toilet use practices, that is, more people in the community are perceived to use the toilet compared to actual prevalence, then conformity with such beliefs may induce an individual to use toilet (Kuang et al. 2020). This overestimation bias or false positive beliefs in perceived prevalence has been found to influence behaviors like exclusive breastfeeding (Bicchieri et al. 2022). Similarly, perception about approval of others about toilet usage may overshoot the actual approval within the community. This false positive belief about approval can also potentially influence behavior. Given the significant positive association of EE and NE with *jati* fractionalization, we also assess whether households in fractionalized areas have false positive beliefs about their community in terms of prevalence of toilet usage and its approval. Additionally, we observe a significantly positive likelihood of overestimation of prevalence and approval rates of toilet usage among respondents residing in *jati* fractionalized regions (Table 4 Columns 4 and 6). Please note that through these two measures, we compare the perception with actual prevalence and approval in the PSU. Therefore, by construction, PSU level heterogeneities to a large extent would be accounted for in the regression.

To sum up, we find that a household residing in fractionalized areas is more likely on average to have positive normative beliefs about toilet usage, higher EE and NE and also false positive beliefs about prevalence and approval of toilet usage within the community. To understand the relevance of these factors in driving systematically higher toilet usage that is observed in fractionalized areas, we assess if there is a change in effect size after controlling for these factors separately in the regression. We observe very limited change in the effect size when regressed with the indicator of PNB compare to that when regressed without it (Figure 1). This underscores the negligible effect of PNB in driving toilet usage and ownership in *jati* diverse areas. Second, we consider EE as mediating factor and observe the changes in effect size when the outcome variables are regressed with and without it. Figure 1, which presents these estimates, shows a discernible drop in the effect of *jati* fractionalization by about 50-58 percentage points. However, when NE is considered as a mediating channel, we do not observe any fall in the effect size, underscoring the limited role of NE. Notably, when both EE and NE are considered as the channels, a fall by 49-57 percentage points is observed. These estimates highlight that social expectation is a key through which *jati* diversity affects improved toilet usage behavior and ownership that we observe with EE playing a significant role though no such implication of NE or PNB is observed. Importantly, we repeat this analysis considering EE and NE bias as potential mediating channels and found a similar inference. The effect on *jati* fractionalization on toilet usage and ownership reduces substantially once the EE bias in toilet usage is controlled for. These results highlight that in *jati* diverse areas, beliefs surrounding prevalence of toilet usage and ownership is higher. These beliefs are on average positively biased and hence higher than the actual prevalence within the community. This creates an important pathway for driving up toilet use and ownership in these regions by creating a descriptive norm (Burztyn et al. 2020). This observation is also consistent with previous studies which document that toilet use behavior in India is higher among individuals with higher EE surrounding toilet usage (Thulin et al., 2021).

*Social learning and Networks*

An obvious question that follows is: why is it that in *jati* diverse region, EE is higher? Why do we observe false positive beliefs on toilet usage or people overestimate prevalence of toilet usage in these diverse regions? One way through which caste diversity can lead to higher prevalence of pro-social collective behavior is the social learning effect wherein people of diverse *jatis* learn about different attitudes, beliefs, and behaviors from each other. In diverse areas, exposure to different perspectives and behavior can play an influential role in shaping up beliefs and attitudes (Gisselquist, 2016; Dinku et al. 2019). If these behaviors are pro-social and progressive, that can lead to positive consequences on the society as a whole.

A very critical aspect of an individual's social expectation and learning is her reference network or people whose behavior or opinion matter to her. Social learning happens when people exchange information and ideas or learns about attitudes and behavior from those within their interpersonal networks. If individuals in one's reference network behave progressively and have pro-social attitudes, it might result in changes in her social expectations, beliefs and thereby in behavior. Therefore, we argue that one potential pathway for the prevalence of higher EE in diverse localities is positive social learning, which originates from greater social interactions among people of diverse *jatis* and learning about their toilet usage behavior and beliefs.

We utilize an egocentric network survey to explore the role of individuals' network in shaping preferences and social expectations on toilet usage. The benefit of egocentric network data is that it allows us to map the connections of the respondent (alters) and gather information from them. In the survey, respondents are asked to name those:

(i) whom they would contact in the event of various crises, such as crop failure, water shortage or a flood

(ii) whom they respect the most in their community

(iii) whom they are more likely to have conversations with about toilet usage practices and open defecation

(iv) whom they are more likely to ask for help or advice if they need to build a latrine

(v) whom they are more likely to ask for help or advice to clean or repair a latrine

These questions, known as "name generators," allow us to see a broad variety of relationships that are prevalent in the surveyed communities. Here we asked to give at most five names for each of the following though the same name can feature in multiple questions. Overall, we collect information about alters from 1501 different egos. Next, we ask specific details from the respondents about each of the alters generated by them about where they stay and their caste along with questions related to toilet usage that includes:

(i) Their (alters') usual place of defecation
(ii) The associated reasons for the alters' toilet usage behavior as reported by the respondent
(iii) Alters' beliefs about OD reported by respondent
(iv) If the respondent had any conversations with them (alters) on toilet usage
(v) If the respondent had any conversations with alters on available models of toilets, availability of masons, cost of construction of toilet, financing it and assistance from Non-Government Organization (NGO) and the government.

From the obtained information on the characteristics of the alters, we look at those who live in the same locality as the ego (respondents). More specifically, we study the nature of interaction, beliefs and behavior surrounding toilet construction and its usage (assessed through the questions above) among the alters staying in the same locality as the ego and then compare those from fractionalized communities and non-fractionalized ones. In particular, we use the share of alters, who use toilets for defecation or with whom the respondent had conversation about toilet usage and construction, among those staying in the same locality as the outcome variables.[13] Table 5 presents the regression estimates of these variables on the fractionalization index with other controls incorporated in the model as used in earlier regressions.

---

[13] See Appendix A for details about the variables.

We observe that the alters from *jati* diverse localities are on average more likely to use toilets for defecation (as reported by the egos) in comparison to those from more homogenous areas. They are also more likely to believe that that one should use toilets because of reasons pertaining to health that include pollution or spread of diseases. In terms of conversations, we find that alters from fractionalized localities are more likely to be interacted with on the type of toilet models or availability of masons though necessarily not on toilet usage or financing toilet construction and assistance received from NGOs or government. Nevertheless, putting all these findings together, we can provide fairly suggestive evidence to infer about improved sanitation behavior and beliefs on toilet usage among residents in jati diverse areas and conversations among them. This underscores the prevalence of progressive social learning around toilet usage among neighbors with fractionalized areas, which also can be linked with higher EE surrounding toilet usage and hence improved sanitation behavior.

There is an important issue one may raise when we present these findings.[14] Even when we document about progressive social learning, a compelling concern might be whether the networks within the locality in *jati* diverse areas are formed among people of same caste. In other words, an individual from these areas might only form networks and interact with those from the same caste. In that case, our argument about social learning from inter-caste interactions might fall apart. To answer this, we use the same ego-centric survey and assess whether the proportion of alters from the same locality and belonging to the same social group as that of the ego is lower from fractionalized area as compared to the non-fractionalized ones. The regression estimates indicate the share of such alters from same caste who are interacted with is significantly lower if the ego is from jati fractionalized areas (Table 5- last column). This alleviates our concern on absence of inter-caste interaction in fractionalized areas.

*Alternative explanations*

---

[14] There is one more issue one may raise on sanctions. If prevalence of toilet usage and conversations around it is higher in diverse areas, is it the case that people here are more likely to be sanctioned for defecating in the open? In the main survey, we ask the respondents about the possibility of doing anything in response if somebody from their locality is found to be defecating in the open. We find that negative actions that include gossips, scolding or fines for someone found openly defecating are more likely to be in place in these fractionalized communities. This assures us that people from fractionalized areas potentially look at open defecation as a health hazard and places sanctions if someone deviates from the norm of using toilets for defecation

*Presence of Muslims*

In the context of Northern India, Muslims on average are more likely to use toilets for defecation even after accounting for their socio-economic characteristics. In a paper by Spears and Geruso (2018), it is argued that Muslim children are more likely to survive their first birth than Hindus because of the externalities derived by the latter due to poor sanitation. In our context, it is likely that the fragmented areas have higher proportion of Muslims which may drive up the average toilet ownership and usage in these communities. Therefore, to account for the presence of Muslims, we generate PSU level of proportion of Muslims and use them as an additional covariate in the main regression. We find very negligible change in the effect size with and without controlling for the proportion of Muslim households (Table 6). This undermines the importance of the presence of Muslims as contributing factor that can explain improved sanitation practices in *jati* diverse areas.

*Sanitation interventions*

In the previous sections, we argue how interactions among diverse communities in fragmented areas can be influential in creating a progressive norm of toilet usage, thereby improving individual sanitation behavior as well. Nevertheless, one may still argue that other sanitation interventions that might be disproportionately predominant in *jati* diverse areas could have led to creating the descriptive norm of toilet usage thereby increasing toilet usage within such communities. Existing literature indicates that the political institutions can create incentives for political parties to work across caste lines in socially diverse areas (Gibson and Hoffman, 2013). This is especially relevant in the context of the prevalent political system in Bihar where caste is a salient factor and a number of political parties within the state are based on caste lines (Jafferlot, 2000; Kumar, 2013). Therefore, it is likely that the volume of interventions on toilet construction, community toilets and interventions to arrest open defecation is higher in these diverse regions. If this is true, then it is also likely that the improved sanitation practices that we observe are driven by the disproportionately higher supply-side interventions in *jati* diverse regions. For example, caste diversity can influence toilet usage through higher provisioning of community public toilets or through government intervention for promoting toilet use. In former case, local political authority in power may provision higher public toilets in caste diverse region to create a vote base among caste groups, thereby increasing toilet usage. In latter case, the government may systematically introduce policies or interventions to promote toilet construction and usage in more diverse regions again to gain political support across caste lines, as a result of which toilet usage and construction may increase in diverse communities.

To test this line of argument, we consider a number of measures that give indications about the measures of government interventions: whether the local schools teach about importance of toilet usage; whether the respondent knows received any interventions on open defecation in the last five years; whether there are any interventions currently implemented locally; whether there is a community toilet near the residence of the respondent; whether the respondent participated in SBA initiatives and whether he/ she has seen any advertisement on SBA. Next, we control for these variables along other covariates and then run the regression with toilet ownership and usage as the outcome variables and *jati* fractionalization as the main variable of interest. The findings indicate no discernible changes in the estimates with and without these variables. We also use participation of political meetings and rallies as an additional covariate but the estimate remain very similar (Figure 2). This underscores the limited role of supply side interventions in explaining improved sanitation practices in caste diverse areas and allows us to negate it as a possible channel.

*Presence of Self-Help Group (SHG)*

On similar lines, one may argue that SHG activities can be disproportionately higher in diverse areas. Studies based in India have found substantial impact of the SHGs on household welfare and women empowerment, which can also lead to improvement in sanitation practices (Deininger and Liu, 2009; Swain and Wallentin, 2009; Raghunathan et al. 2022). To test the presence of SHG as a potential channel, we use participation in SHGs as an additional control variable in the main regressions. We find very negligible changes in the effect size after its introduction as covariate when compared to that without the variable. Therefore, we are also able to negate presence of SHGs as a potential mechanism explaining the diversity dividend finding that we observe (Figure 3).

## 6. Conclusion

Existing literature have identified the impact of caste diversity on education, health, and provisioning of public goods and have been largely equivocal about the loss, which higher diversity places on the residents and the society. While studies have also documented diversity gains, not much emphasis has been given on understanding the implications that higher diversity may carry. In this paper, we first assess the effects of local level sub-caste (*jati*) diversity on toilet access and usage behavior. Second, we examine the role of social expectations and beliefs and then explore the relevant social network characteristics to investigate the importance of social learning in explaining the effects of *jati* diversity. We use survey data collected from eight districts of Bihar during 2017-2018 to measure sanitation behavior.

The findings indicate that toilet usage and access is discernibly higher in *jati* diverse regions. We also observe the significant role played by social expectations wherein perception about community level prevalence of toilet usage (EE) is found to be higher on average for respondents from fractionalized communities. Borrowing from the conceptual framework of SNT, we find that higher EE about toilet usage plays a mediating role that improves sanitation behavior in fractionalized areas. Next, to understand the reasons for higher EE in fractionalized communities, we use a longitudinal ego-centric network survey conducted six months prior to the main survey to assess the role of social networks in shaping up the preferences and social expectations. We find that interactions with neighbors residing in diverse communities often pertain to discussion about toilet models and availability of masons. Importantly, these neighbors, who the respondents from diverse localities interact with, are likely to use toilet for defecation and believe that OD can increase pollution and spread diseases. This provides compelling evidence of the importance of higher social leaning leading to improved sanitation practices in these areas.

The findings of this study hold significant policy implications for promoting progressive and pro-social behavior that include toilet usage and access. Focus on ways to foster diversity and social mixing within communities and encouraging diverse community interactions through community programs that bring together individuals from different backgrounds can be emphasized. Furthermore, the paper underscores the importance of localized approaches. Policies that can target specific neighborhoods or regions with high diversity to implement tailored interventions that leverage social learning mechanisms may be encouraged. These interventions might include awareness campaigns, descriptive norm messaging, community workshops, or peer-to-peer initiatives that facilitate knowledge sharing and behavior change related to toilet usage. Additionally, financial assistance programs or subsidies targeted in these areas can help overcome any economic barriers to toilet construction and maintenance. In summary, the policy implications suggest that fostering diversity and promoting social interactions in communities can play a pivotal role in promoting improved toilet usage and access. By recognizing the influence of social learning and tailoring interventions to localized contexts, policymakers can effectively address the sanitation challenges owing to behavioral factors, ultimately leading to better public health outcomes and improved quality of life.

# References

Allport, G.W., K. Clark, and T. Pettigrew (1954). The Nature of Prejudice. New York: Addison-Wesley.

Alesina, A., Baqir, R., & Easterly, W. (1999). Public goods and ethnic divisions. *The Quarterly Journal of Economics*, *114*(4), 1243-1284.

Altonji, J. G., Elder, T. E., & Taber, C. R. (2005). Selection on observed and unobserved variables: Assessing the effectiveness of Catholic schools. *Journal of Political Economy*, *113*(1), 151-184.

Azar, J., Marinescu, I., & Steinbaum, M. (2022). Labor market concentration. *Journal of Human Resources*, *57*(S), S167-S199.

Baker, K. K., B. Padhi, B. Torondel, P. Das, A. Dutta, K. C. Sahoo, B. Das, et al. 2017. "From Menarche to Menopause: A Population-Based Assessment of Water, Sanitation, and Hygiene Risk Factors for Reproductive Tract Infection Symptoms Over Life Stages in Rural Girls and Women in India." PLoS One 12 (12): e0188234.

Baker, K. K., W. T. Story, E. Walser-Kuntz, and M. B. Zimmerman. 2018. "Impact of Social Capital, Harassment of Women and Girls, and Water and Sanitation Access on Premature Birth and low Infant Birth Weight in India." PLoS One 13 (10): e0205345.

Banerjee, A. N., Banik, N., & Dalmia, A. (2017). Demand for household sanitation in India using NFHS-3 data. *Empirical Economics*, *53*(1), 307-327.

Balasubramaniam, D., Chatterjee, S., & Mustard, D. B. (2014). Got water? Social divisions and access to public goods in rural India. *Economica*, 81(321), 140-160.

Bevis, L., Kim, K., & Guerena, D. (2023). Soil zinc deficiency and child stunting: Evidence from Nepal. *Journal of Health Economics*, *87*, 102691.

Bicchieri, C. (2005). *The grammar of society: The nature and dynamics of social norms*. Cambridge University Press.

Bicchieri, C. (2016). *Norms in the wild: How to diagnose, measure, and change social norms*. Oxford University Press.

Bicchieri, C., Das, U., Gant, S., & Sander, R. (2022). Examining norms and social expectations surrounding exclusive breastfeeding: Evidence from Mali. *World Development*, *153*, 105824.

Biswas, S., & Das, U. (2022). Adding fuel to human capital: Exploring the educational effects of cooking fuel choice from rural India. *Energy Economics*, *105*, 105744.

Bursztyn, L., González, A. L., & Yanagizawa-Drott, D. (2020). Misperceived social norms: Women working outside the home in Saudi Arabia. *American Economic Review*, *110*(10), 2997-3029.

Cameron, L. A., Shah, M., & Olivia, S. (2013). Impact evaluation of a large-scale rural sanitation project in Indonesia. *World Bank policy research working paper*, (6360).

Carrasco, J. A., Hogan, B., Wellman, B., & Miller, E. J. (2008). Collecting social network data to study social activity-travel behavior: an egocentric approach. *Environment and Planning B: Planning and Design*, *35*(6), 961-980.

Cao, B., Saffer, A. J., Yang, C., Chen, H., Peng, K., Pan, S. W., ... & Tucker, J. D. (2019). MSM behavior disclosure networks and HIV testing: an egocentric network analysis among MSM in China. *AIDS and Behavior*, *23*, 1368-1374.

Churchill, S. A., Farrell, L., & Smyth, R. (2019). Neighbourhood ethnic diversity, and mental health in Australia. *Health Economics*, 28(9), 1075-1087.

Churchill, S. A., & Smyth, R. (2020). Ethnic diversity, emergy poverty, and the mediating role of trust: Evidence from household panel for Australia. *Energy Economics*, 86, 104663.

Conley, T. G., Hansen, C. B., & Rossi, P. E. (2012). Plausibly exogenous. *Review of Economics and Statistics*, *94*(1), 260-272.

Deininger, K., & Liu, Y. (2009). Economic and social impacts of self-help groups in India. *World Bank Policy Research Working Paper*, (4884).

Desai, S., & Dubey, A. (2011). Caste in 21st century India: Competing narratives. *Economic and Political Weekly*, 40-49.

Deshpande, A. (2000). Does caste still define disparity? A look at inequality in Kerala, India. *American Economic Review*, *90*(2), 322-325.

Deshpande, A. (2011). *The grammar of caste: Economic discrimination in contemporary India*. Oxford University Press.

Dinku, Y., Fielding, D., & Genç, M. 2019. Neighbourhood ethnic diversity, child health outcomes and women's empowerment. *The Journal of Development Studies*, 55(9), 1909-1927.

Dumont, L. (1980). *Homo hierarchicus: The caste system and its implications*. University of Chicago Press.

Easterly, W., & Levine, R. (1997). Africa's growth tragedy: policies and ethnic divisions. *Quarterly Journal of Economics*, 1203-1250.

Finseraas, H., Hanson, T., Johnsen, Å. A., Kotsadam, A., & Torsvik, G. (2019). Trust, ethnic diversity, and personal contact: A field experiment. *Journal of Public Economics*, *173*, 72-84.

Deutsch, M. & Gerard, H.B. (1955). A study of normative and informational social influences upon individual judgment. *Journal of Abnormal and Social Psychology,* 51, 629-636

Gibson, C. C., & Hoffman, B. D. (2013). Coalitions not conflicts: Ethnicity, political institutions, and expenditure in Africa. *Comparative Politics*, *45*(3), 273-290.

Gisselquist, R. M., Leiderer, S., & Nino-Zarazua, M. (2016). Ethnic heterogeneity and public goods provision in Zambia: Evidence of a subnational "diversity dividend". *World Development*, 78, 308-323.

Harter, M., Contzen, N., & Inauen, J. (2019). The role of social identification for achieving an open-defecation free environment: A cluster-randomized, controlled trial of Community-Led Total Sanitation in Ghana. *Journal of environmental psychology*, *66*, 101360.

Hutton, G., Rodriguez, U. E., Napitupulu, L., Thang, P., & Kov, P. (2007). Economic impacts of sanitation in southeast Asia: summary. *World Bank Policy Research Working Paper*, (44121).

Jaffrelot, C. (2000). The rise of the other backward classes in the Hindi belt. *The Journal of Asian Studies*, *59*(1), 86-108.

Kruglanski, A. W. (1989). The psychology of being" right": The problem of accuracy in social perception and cognition. *Psychological Bulletin*, *106*(3), 395.

Khwaja, A. I. (2009). Can good projects succeed in bad communities? *Journal of Public Economics*. 93 (7-8), 899-916.

Kuang, J., Thulin, E., Ashraf, S., Shpenev, A., Das, U., Delea, M. G., ... & Bicchieri, C. (2020). Bias in the perceived prevalence of open defecation: Evidence from Bihar, India. *PloS one*, *15*(9), e0238627.

Kumar, A. (2013). Nitish Kumar's Honourable Exit: A brief history of caste politics. *Economic and Political Weekly*, 15-17.

Kumar, S., & Vollmer, S. (2013). Does access to improved sanitation reduce childhood diarrhea in rural India? *Health Economics*, *22*(4), 410-427.

McArthur, J. W., & McCord, G. C. (2017). Fertilizing growth: Agricultural inputs and their effects in economic development. *Journal of Development Economics*, *127*, 133-152.

Miguel, E., & Gugerty, M. K. (2005). Ethnic diversity, social sanctions, and public goods in Kenya. *Journal of Public Economics*, *89*(11-12), 2325-2368.

Montalvo, J. G., & Reynal-Querol, M. (2021). Ethnic diversity and growth: Revisiting the evidence. Review of Economics and Statistics, 103(3), 521-532.

Munshi, K. (2019). Caste and the Indian economy. *Journal of Economic Literature*, *57*(4), 781-834.

O'malley, A. J., Arbesman, S., Steiger, D. M., Fowler, J. H., & Christakis, N. A. (2012). Egocentric social network structure, health, and pro-social behaviors in a national panel study of Americans. *PloS One*, *7*(5), e36250.

Oster, E. (2014). Unobservable selection and coefficient stability. *Chicago: University of Chicago Booth School of Business*.

Raghunathan, K., Kumar, N., Gupta, S., Thai, G., Scott, S., Choudhury, A., ... & Quisumbing, A. (2022). Scale and sustainability: The impact of a women's self-help group program on household economic well-being in India. *The Journal of Development Studies*, 1-26.

Rathore, U., & Das, U. (2022). Health consequences of patriarchal kinship system for the elderly: evidence from India. *The Journal of Development Studies*, *58*(1), 145-163.

Reid, A. E., Cialdini, R. B., & Aiken, L. S. (2010). Handbook of behavioral medicine: Methods and applications, Springer New York, New York, NY.

Scacco, A., & Warren, S. S. (2018). Can social contact reduce prejudice and discrimination? Evidence from a field experiment in Nigeria. *American Political Science Review*, *112*(3), 654-677.

Sclar, G. D., Garn, J. V., Penakalapati, G., Alexander, K. T., Krauss, J., Freeman, M. C., & Clasen, T. (2017). Effects of sanitation on cognitive development and school absence: a systematic review. *International Journal of Hygiene and Environmental Health*, *220*(6), 917-927.

Sclar, G. D., G. Penakalapati, B. A. Caruso, E. A. Rehfuess, J. V. Garn, K. T. Alexander, M. C. Freeman, S. Boisson, K. Medlicott, and T. Clasen. 2018. "Exploring the Relationship Between Sanitation and Mental and Social Well-Being: A Systematic Review and Qualitative Synthesis." Social Science & Medicine 217: 121–134.

Singh 1992. Singh, K. S, ed. 1992. An Introduction. Vol. 36, People of India. Calcutta: Anthropological Survey of India.

Stock, J. H., Wright, J. H., & Yogo, M. (2002). A survey of weak instruments and weak identification in generalized method of moments. *Journal of Business & Economic Statistics*, *20*(4), 518-529.

Sunstein, C. R. (1996). Social norms and social roles. *Columbia Law Review*, *96*, 903.

Suls, J., Wan, C. K., & Sanders, G. S. (1988). False consensus and false uniqueness in estimating the prevalence of health-protective behaviors. *Journal of Applied Social Psychology*, *18*(1), 66-79.

Swain, R. B., & Wallentin, F. Y. (2009). Does microfinance empower women? Evidence from self-help groups in India. *International Review of Applied Economics*, *23*(5), 541-556.

Thulin, E., Shpenev, A., Ashraf, S., Das, U., Kuang, J., & Bicchieri, C. (2021). Toilet Use is a Descriptive Norm: The Influence of Social Expectations on Toilet Use in Bihar and Tamil Nadu, India. *India (October 2021)*.

Wooldridge, J. M. (2010). *Econometric analysis of cross section and panel data*. MIT press.

Vaitla, B., Taylor, A., Van Horn, J., & Cislaghi, B. (2017). Social Norms and Girls' Well-Being– Integrating Theory, Practice, and Research.

Van Minh, H., & Hung, N. V. (2011). Economic aspects of sanitation in developing countries. *Environmental Health Insights*, *5*, EHI-S8199.

**Figures**

**Figure 1. Role of Social Expectations**

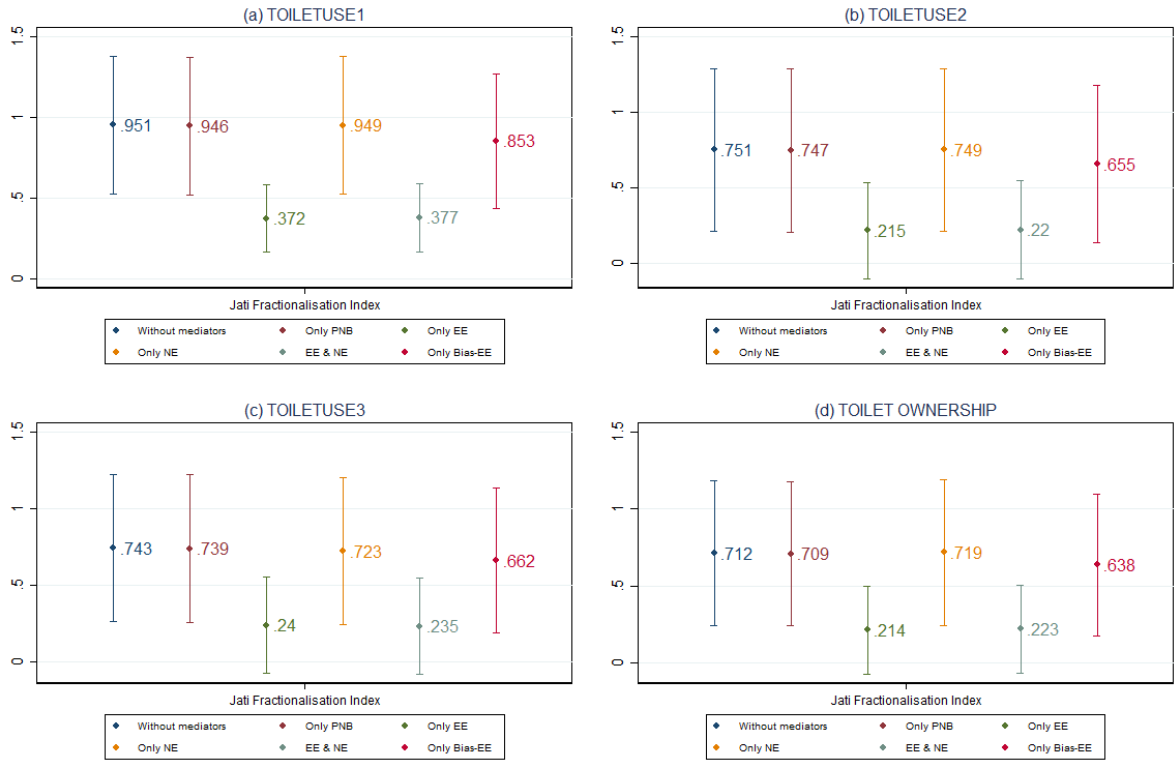

**Note:** Marginal effects from OLS regressions are presented along with 95% Confidence Interval calculated by clustering the standard errors at PSU level. TOILETUSE1 measures toilet use behavior for the last time they defecated; TOILETUSE2 measures the toilet use behavior for the last three time they defecated, TOILETUSE3 measures the toilet use behavior in the last seven day they defecated; Toilet Ownership measures if the household owns the toilet. EE stands for Empirical Expectations, NE for Normative Expectations and PNB stands for Personal Normative Beliefs. The details on control variables is given in Table 1. The details on variable construction is given in Appendix A.

# Figure 2. Sanitation Interventions

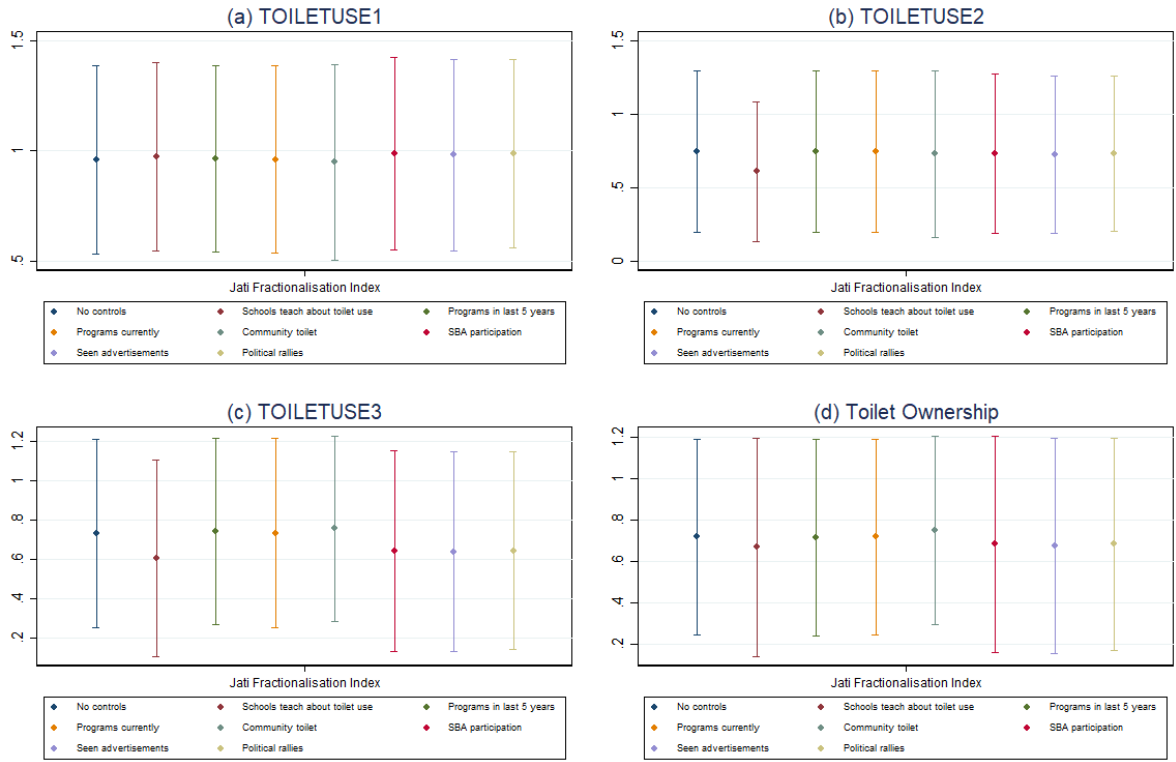

**Note:** Marginal effects from OLS regressions are presented along with 95% Confidence Interval calculated by clustering the standard errors at PSU level. TOILETUSE1 measures toilet use behavior for the last time they defecated; TOILETUSE2 measures the toilet use behavior for the last three time they defecated; TOILETUSE3 measures the toilet use behavior in the last seven day they defecated; Toilet Ownership measures if the household owns the toilet. The details on control variables is given in Table 1. The details on other variables and variable construction is given in Appendix A.

# Figure 3. Presence of Self-Help Groups

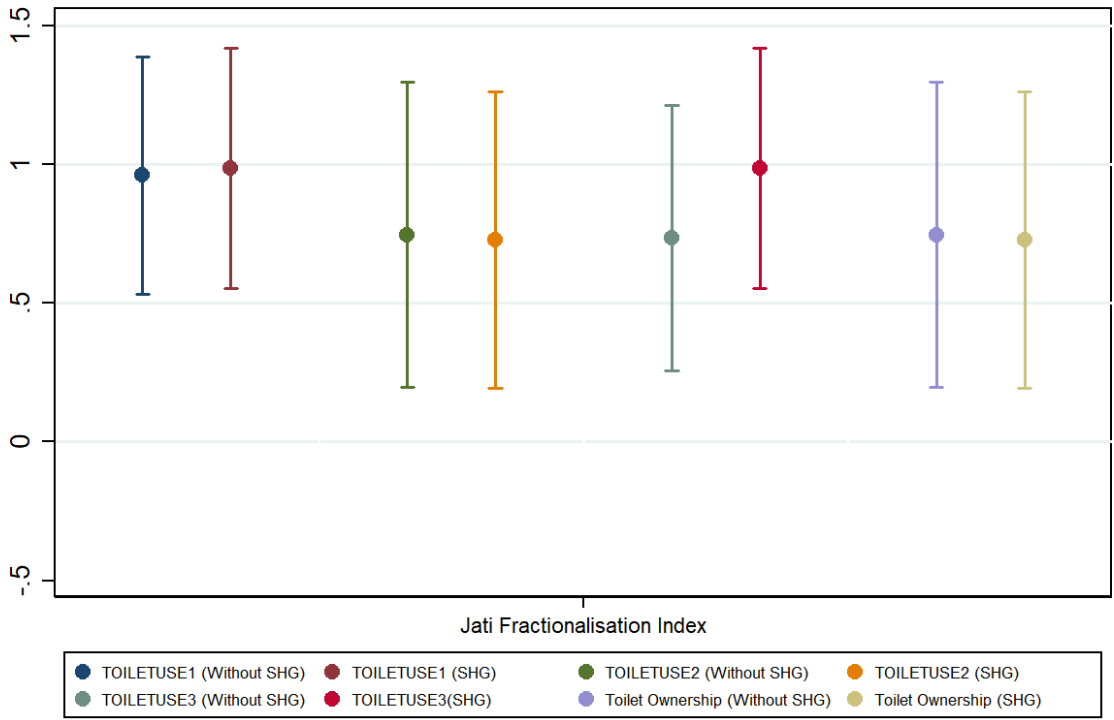

**Note:** Marginal effects from OLS regressions are presented along with 95% Confidence Interval calculated by clustering the standard errors at PSU level. TOILETUSE1 measures toilet use behavior for the last time they defecated; TOILETUSE2 measures the toilet use behavior for the last three time they defecated; TOILETUSE3 measures the toilet use behavior in the last seven day they defecated; Toilet Ownership measures if the household owns the toilet. The details on control variables is given in Table 1. The details on other variables and variable construction is given in Appendix A.

# Tables

## Table 1: Summary Statistics

| Variable | Description | Mean | SD |
|---|---|---|---|
| TOILETUSE1 | One if individual use toilet (public/private) the last time when defecated | 0.592 | 0.492 |
| TOILETUSE2 | One if individual use toilet (public/private) in the last three times when defecated | 0.535 | 0.499 |
| TOILETUSE3 | One if individual use toilet (public/private) in the last week when defecated | 0.476 | 0.500 |
| TOILET OWNERSHIP | One if individual has a toilet in his/her household | 0.587 | 0.492 |
| FRAC | *Jati* fractionalization index | 0.714 | 0.209 |
| AGE | Individual's age in years | 34.927 | 13.820 |
| FEMALE | One if individual is female | 0.519 | 0.500 |
| EDUCATION | Individual's years of education | 5.403 | 5.493 |
| HINDU | One if household is Hindu | 0.737 | 0.440 |
| MUSLIM | One if household is Muslim | 0.254 | 0.436 |
| MOTORCYCLE | One if household has Motorcycle | 0.233 | 0.423 |
| COLOR TV | One if household has Color TV | 0.453 | 0.498 |
| REFRIGERATOR | One if household has Refrigerator | 0.111 | 0.314 |
| RATION CARD | One if household has BPL or *Antodaya* ration card | 0.562 | 0.496 |
| KITCHEN | One if household has separate room for kitchen | 0.308 | 0.462 |
| PHWATER | Proportion of households without water source inside house in the village | 0.229 | 0.114 |
| PHBPL | Proportion of BPL ration card holder households in PSU | 0.562 | 0.157 |
| PHWOMEN | Proportion of literate women in the PSU | 0.449 | 0.152 |
| EE | Empirical Expectation: Proportion of people you think have used toilet the last time they needed to defecate | 0.612 | 0.353 |
| BIAS EE | =1 if an individual's perceived EE on toilet use is higher than community-level actual prevalence of toilet use | 0.555 | 0.497 |
| NE | Normative Expectation: Proportion of people you think believe that it is wrong to defecate in open | 0.853 | 0.242 |
| BIAS NE | =1 if an individual's perceived NE on toilet use is higher than community-level actual prevalence of toilet use | 0.555 | 0.497 |
| PNB | Personal Normative Beliefs. = 1 if an individual personally believes that it is wrong to defecate in the open | 0.959 | 0.198 |
| FB | Factual beliefs. = 1 if an individual personally believes that there are any bad health effects of open defecation for you | 0.874 | 0.332 |
| Observations | | 2533 | |

**Note:** SD stands for standard deviation. The details on variable construction is given in Appendix A.

## Table 2: Toilet Use, Ownership and *Jati* Diversity

|  | TOILETUSE1 | | TOILETUSE2 | | TOILETUSE 3 | | Toilet Ownership | |
|---|---|---|---|---|---|---|---|---|
|  | (1) | (2) | (3) | (4) | (5) | (6) | (7) | (8) |
| *Jati* Fractionalisation | 0.850*** | 0.961*** | 0.663*** | 0.745*** | 0.733*** | 0.622*** | 0.616*** | 0.717*** |
|  | (0.179) | (0.209) | (0.240) | (0.269) | (0.234) | (0.204) | (0.202) | (0.231) |
| Constant | -0.141 | -0.375 | -0.070 | -0.250 | -0.042 | -0.138 | -0.079 | -0.489 |
|  | (0.147) | (0.291) | (0.198) | (0.353) | (0.284) | (0.180) | (0.157) | (0.321) |
| Demographic controls | Yes | Yes | Yes | Yes | Yes | Yes | Yes | Yes |
| PSU controls | No | Yes | No | Yes | Yes | No | No | Yes |
| District FE | Yes | Yes | Yes | Yes | Yes | Yes | Yes | Yes |
| R-squared | 0.415 | 0.433 | 0.357 | 0.368 | 0.320 | 0.307 | 0.371 | 0.391 |
| N | 2533 | 2533 | 2533 | 2533 | 2533 | 2533 | 2533 | 2533 |

**Note:** Marginal effects from OLS regressions are presented. The details on control variables is given in Table 1. The details on variable construction is given in Appendix A. Standard errors are reported in parentheses and clustered at PSU level. ***, **, and * represent significant level at 1%, 5% and 10% respectively.

## Table 3: IV Estimates of *Jati* Diversity on Toilet Use and Ownership

|  | (1) | (2) | (3) | (4) |
|---|---|---|---|---|
|  | IV = Religion-caste Fractionalisation Index | | | |
|  | TOILETUSE1 | TOILETUSE2 | TOILETUSE3 | Toilet Ownership |
| *Jati* Fractionalisation | 0.974*** | 0.740** | 0.541 | 0.752*** |
|  | (0.235) | (0.327) | (0.333) | (0.241) |
| Observations | 2533 | 2533 | 2533 | 2533 |
| R-squared | 0.260 | 0.179 | 0.184 | 0.249 |
| Demographic Characteristics | Yes | Yes | Yes | Yes |
| PSU controls | Yes | Yes | Yes | Yes |
| District Fixed effects | Yes | Yes | Yes | Yes |
| First Stage F-Stat/Kleibergen-Paap Wald rk-F stat (Weak ID) |  |  | 22.17 |  |
| Cragg-Donald Wald F stat (Weak ID) |  |  | 2411.81 |  |
| Kleibergen-Paap rk LM Chi-sq stat (Under ID) |  |  | 8.16 |  |
| Stock-Yogo Weak ID F-test critical values at 5% | 10% | 15% | 20% | 25% |
|  | 16.38 | 8.96 | 6.66 | 5.53 |

**Note:** Marginal Effects from 2SLS IV regression are presented. The details on control variables is given in Table 1. The details on variable construction is given in Appendix A. Standard errors are reported in parentheses and clustered at PSU level. ***, **, and * represent significant level at 1%, 5% and 10% respectively.

### Table 4. Mechanism: Role of Social Expectations

|  | (1) Factual beliefs (FB) | (2) Personal Normative beliefs (PNB) | (3) Empirical Expectation (EE) | (4) =1 if Overestimated Bias (Bias EE) | (5) Normative Expectation (NE) | (6) =1 if Overestimated Bias (Bias NE) |
|---|---|---|---|---|---|---|
| *Jati* Fractionalisation | 0.061 | 0.061** | 0.917*** | 0.533*** | 0.114* | 0.358 |
|  | (0.050) | (0.022) | (0.203) | (0.156) | (0.065) | (0.253) |
| R-squared | 0.064 | 0.038 | 0.542 | 0.115 | 0.051 | 0.146 |
| N | 2533 | 2533 | 2533 | 2533 | 2533 | 2533 |

**Note:** Marginal effects from OLS regressions are presented. Standard errors are reported in parentheses and clustered at PSU level. The details on dependent variables and their construction is given in Appendix A. ***, **, and * represent significant level at 1%, 5% and 10% respective

### Table 5. Mechanism: Social Learning and Networks

|  | (1) Alters' use toilets | (2) Health as reason | (3) Belief on OD is wrong | (4) Converse about toilet use | (5) Toilet models | (6) Masons | (7) Construction cost | (8) Financing | (9) Help from NGO | (10) Help from Govt | (11) Alters' from same caste |
|---|---|---|---|---|---|---|---|---|---|---|---|
| *Jati* Fractionalisation | 0.734*** | 0.378** | -0.120 | 0.110 | 0.129** | 0.115* | 0.081 | -0.133 | 0.042 | -0.214 | -0.357** |
|  | (0.096) | (0.067) | (0.090) | (0.084) | (0.050) | (0.061) | (0.066) | (0.104) | (0.038) | (0.160) | (0.154) |
| R-squared | 0.194 | 0.091 | 0.091 | 0.054 | 0.037 | 0.038 | 0.038 | 0.129 | 0.038 | 0.142 | 0.075 |
| N | 923 | 923 | 923 | 923 | 923 | 923 | 923 | 923 | 923 | 923 | 923 |

**Note:** Marginal effects from OLS regressions are presented. The description of dependent variables and their construction is given in Appendix A Standard errors are reported in parentheses and clustered at PSU level. ***, **, and * represent significant level at 1%, 5% and 10% respectively.

## Table 6. Presence of Muslims

|  | TOILETUSE1 | | TOILETUSE2 | | TOILETUSE3 | | Toilet Ownership | |
|---|---|---|---|---|---|---|---|---|
|  | (1) | (2) | (3) | (4) | (5) | (6) | (7) | (8) |
| *Jati* Fractionalisation | 0.961*** | 0.931*** | 0.745*** | 0.712*** | 0.733*** | 0.761*** | 0.717*** | 0.750*** |
|  | (0.209) | (0.202) | (0.269) | (0.250) | (0.234) | (0.245) | (0.231) | (0.236) |
| Proportion of muslims in PSU |  | 0.099 |  | 0.111 |  | -0.092 |  | -0.107 |
|  |  | (0.112) |  | (0.133) |  | (0.147) |  | (0.120) |
| R-squared | 0.433 | 0.433 | 0.368 | 0.369 | 0.320 | 0.321 | 0.391 | 0.392 |
| N | 2533 | 2533 | 2533 | 2533 | 2533 | 2533 | 2533 | 2533 |

**Note:** Marginal effects from OLS regressions are presented. The details on variable construction is given in Appendix A. Standard errors are reported in parentheses and clustered at PSU level. ***, **, and * represent significant level at 1%, 5% and 10% respectively.

**Appendix A. Description and Construction of variables**

### A. Dependent variables in Table 1

**TOILETUSE1:** Some people use a latrine, and some people defecate in the open. Where did you defecate the last time you had to? *a. Used my household's private latrine; b. Household, c. Used a community/public toilet, d. Defecated in the open*

Dummy variable: If the answer is (a)-(c), then coded as 1. If (d), then coded as 0.

**TOILETUSE2:** In the last three times you defecated, how many of those were in the open?

Dummy variable: If the answer is 3, then coded as 1. If answer is 0, 1 or 2, the coded as 0.

**TOILETUSE3:** In the past week, how often have you used a latrine to defecate? Never, occasionally, frequently, or every time? *a. Never; b. Occasionally; c. Frequently; d. Every time*

Dummy variable: If the answer is (d), then coded as 1. If answer is (a)-(c), then coded as 0.

**TOILET OWNERSHIP:** Does your household own a latrine? *a. No latrine; b. Sole owner; c. Shared with other household*

Dummy variable: If the answer is (c), then coded as 1. If answer is (a)-(b), then coded as 0.

### B. Dependent variables in Table 3

**Factual Beliefs:** Do you personally believe there are any bad health effects of open defecation for you? *a. Yes; b. No*

Dummy variable: If the answer is (a), then coded as 1. If answer is (b), then coded as 0.

**Personal Normative Beliefs (PNB):** Society may think it is right or wrong to defecate in the open. Do you personally think it is right, neither right nor wrong, or wrong for someone to defecate in the open? *a. Yes; b. No*

Dummy variable: If the answer is (a), then coded as 1. If answer is (b), then coded as 0.

**Empirical Expectations (EE):** Out of ten members of your community, how many do you think used a latrine the last time they needed to defecate? *(Numerical value between 0 and 10)*

Numerical value between 0 and 1: Linearly transform between 0 and 1 by dividing with 10.

**Normative Expectations (NE):** Out of ten members of your community, how many do you think believe that it is wrong to defecate in the open? *(Numerical value between 0 and 10)*

Numerical value between 0 and 1: Linearly transform between 0 and 1 by dividing with 10.

**BIAS EE**

Dummy variable: It takes value 1 if individual EE is higher than community-level actual prevalence of toilet use and 0 otherwise. To calculate community-level actual prevalence of toilet use, we take average of TOILETUSE1 at PSU level.

**BIAS NE**

Dummy variable: It takes value 1 if individual NE is higher than community-level beliefs on toilet use and 0 otherwise. To calculate community-level beliefs on toilet use, we take average of personal normative beilefs (PNB) at PSU level.

### C. Dependent variables in Table 5

**1. Alters' use toilets:** Some people use latrines and some people defecate in the open. Does [NAME] usually defecate in the open or use a latrine? *a. Defecate in the open; b. Use a community toilet; c. Use a public toilet; d. Use a household toilet*

**2. Health as reason:** Given the [NAME] uses toilet for defecation, what reasons do you think [NAME] has for this? *a. Prestige/Convience; b. Social Pressure; c. Health and Pollution reasons; d. Not Willful/Other*

**3. Belief on OD:** Does [NAME] think that it is wrong for people to defecate in the open? *a. Yes; b. No*

**4. Converse about toilet use:** Have you had conversations with [NAME] about why people might want to use latrines or defecate in the open in the last 12 months? *a. Yes; b. No*

**5. Toilet models:** Have you ever talked to [NAME] about the available models of toilets? *a. Yes; b. No*

**6. Masons:** Have you ever talked to [NAME] about the availability of masons? *a. Yes; b. No*

**7. Construction cost:** Have you ever talked to [NAME] about how much it costs to construct a toilet? *a. Yes; b. No*

**8. Financing:** Have you ever talked to [NAME] about ways of financing toilet construction? *a. Yes; b. No*

**9. Help from NGO:** Have you ever talked to [NAME] about a non-governmental organization that can potentially help in building a toilet? *a. Yes; b. No*

**10. Help from Govt.:** Have you ever talked to [NAME] about seeking out government help in building a toilet? *a. Yes; b. No*

All of these variables are in proportions as a numerical value between 0 and 1. We first calculate the total number of alters' (networks) who co-reside with respondent in the same neighborhood.

Next, for variable (1), we calculate the total number of alters' (co-residing in the same neighborhood) who use any toilet for defecation.

For variable (2), we calculate the total number of alters' (co-residing in the same neighborhood) who uses toilet because of health and pollution reasons.

For variables (3)-(10), we calculate total number of alters' (co-residing in the same neighborhood) with whom the respondent had any conversations regarding respective topic.

In the last step, we divide these variables with total number of alters' (networks) who co-reside with respondent in the same neighborhood, to calculate the proportions. These variables are calculated only for egos having atleast one alters in their neighborhood.

### D. Additional controls in Figure 3 and Figure 4

**Schools teach about toilet use:** Does the local school teach children to use latrines?

**Programs in last 5 years:** In the last 5 years, have there been any programs in your area to stop open defecation?

**Programs currently:** Are there currently any programs in your area to stop open defecation?

**Community toilet:** Is there a community or a public toilet within 15 minutes of walking from where you live?

**SBA participation:** Have you ever participated in Swachh Bharat Abhiyan campaigns to clean up the area/ village/ city?

**Seen advertisements:** Have you ever seen an advertisement that promotes latrines?

**Political rallies:** Do you participate in campaigns or rallies for any political party?

**SHG:** Are you a member of a village self-help group?

All the above questions have been constructed into dummy variable which takes value 1 if the response is yes and 0 if response is No.

**Appendix B**

**Figure B1: Frequency Distribution of *Jati* Fractionalisation Index**

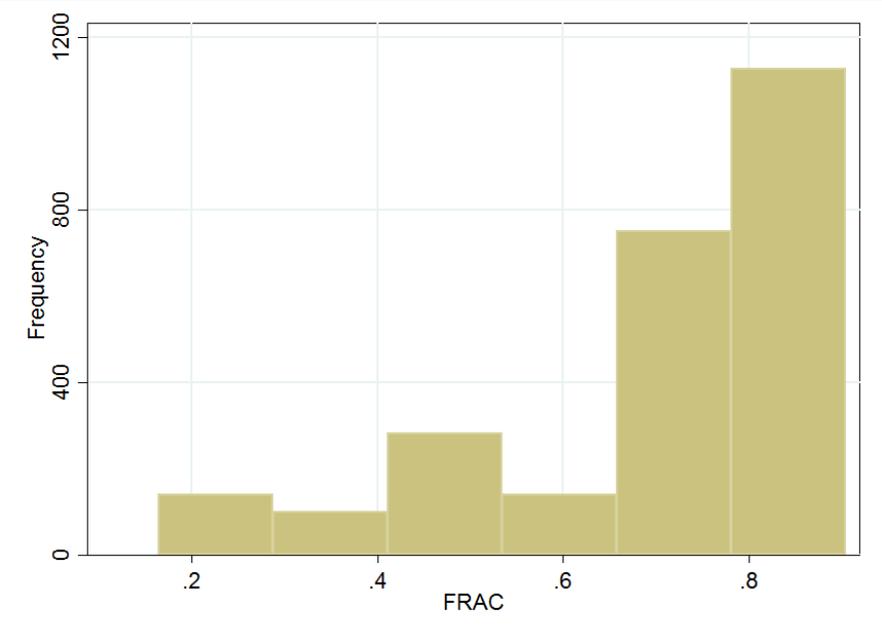

# Tables

**Table B1: Correlation of IV with outcome variables and First-stage estimates**

|  | (1) TOILETUSE1 | (2) TOILETUSE2 | (3) TOILETUSE3 | (4) Toilet Ownership | (5) Jati Fractionalisation |
|---|---|---|---|---|---|
| *Jati* Fractionalisation | 0.948*** | 0.750** | 0.918*** | 0.684** |  |
|  | (0.293) | (0.348) | (0.321) | (0.288) |  |
| Religion-caste Fractionalisation | 0.017 | -0.006 | -0.248 | 0.044 | 0.658*** |
|  | (0.222) | (0.267) | (0.245) | (0.186) | (0.140) |
| R-squared | 0.433 | 0.368 | 0.322 | 0.391 | 0.864 |
| N | 2533 | 2533 | 2533 | 2533 | 2533 |

**Note:** Marginal effects from OLS regressions are presented. The description of dependent variables and their construction is given in Appendix A. Standard errors are reported in parentheses and clustered at PSU level. ***, **, and * represent significant level at 1%, 5% and 10% respectively.

**Table B2: Plausibly exogenous estimates**

|  | (1) TOILETUSE1 | (2) Toilet Ownership |
|---|---|---|
| $\hat{\gamma}$ | 0.641** | 0.495 |
|  | (0.255) | (0.221) |
| Demographic Controls | Y | Y |
| District FE | Y | Y |
| Observations | 2533 | 2533 |
| R-squared | 0.411 | 0.380 |
| $\hat{\beta}$ (UB) | 1.164 | 0.950 |
| $\hat{\beta}$ (LB) | -0.194 | -0.193 |
| $\gamma_{max}$ | 0.51 | 0.36 |

Note: ***, **, and * represent significant level at 1%, 5% and 10% respectively.

### Table B3: Probit Estimates and Alternative measures of Toilet Access

|  | (1) | (2) | (3) | (4) | (5) | (6) |
|---|---|---|---|---|---|---|
|  | PROBIT Estimates | | | | OLS estimates | |
|  | ToiletUse1 | ToiletUse2 | ToiletUse3 | Toilet Ownership | Toilet functional | Toilet access |
| *Jati* Fractionalisation | 3.020*** | 2.141** | 2.177*** | 2.242*** | 0.723*** | 0.801*** |
|  | (0.819) | (0.924) | (0.756) | (0.847) | (0.233) | (0.242) |
| Constant | -3.767*** | -3.012*** | -2.326*** | -4.255*** | -0.525 | -0.703** |
|  | (0.916) | (1.088) | (0.840) | (0.952) | (0.314) | (0.331) |
| R-squared |  |  |  |  | 0.399 | 0.401 |
| N | 2533 | 2533 | 2533 | 2533 | 2533 | 2533 |

**Note:** The description of dependent variables and their construction is given in Appendix A. Standard errors are reported in parentheses and clustered at PSU level. ***, **, and * represent significant level at 1%, 5% and 10% respectively.

### Table B4: Wild Bootstrap Inference

|  | (1) | (2) |
|---|---|---|
|  | T-statistic | P-value |
| **Dependent Variable** |  |  |
| ToiletUse1 | 4.599 | 0.003 |
| ToiletUse2 | 2.774 | 0.073 |
| ToiletUse3 | 3.138 | 0.027 |
| Toilet Ownership | 3.108 | 0.054 |

**Note:** The T-statistic and P-value are calculated from Roodman et al. (2019) methodology.

### Table B5: Census Fractionalisation Index

|  | (1) | (2) | (3) | (4) | (5) |
|---|---|---|---|---|---|
|  | TOILETUSE1 | TOILETUSE2 | TOILETUSE3 | Toilet Ownership | *Jati* Fractionalisation |
| Fractionalisation Index from census | 0.235** | 0.012 | 0.138 | 0.330*** | 0.386*** |
|  | (0.098) | (0.104) | (0.107) | (0.100) | (0.027) |
| R-squared | 0.389 | 0.342 | 0.295 | 0.369 | 0.754 |
| N | 2533 | 2533 | 2533 | 2533 | 2533 |

**Note:** Marginal effects from OLS regressions are presented. The description of dependent variables and their construction is given in Appendix A. Standard errors are reported in parentheses and clustered at PSU level. ***, **, and * represent significant level at 1%, 5% and 10% respectively.